# Inflation of 430-parsec bipolar radio bubbles in the Galactic Centre by an energetic event

I. Heywood[1,2,3]*, F. Camilo[3]*, W. D. Cotton[4,3], F. Yusef-Zadeh[5], T. D. Abbott[3], R. M. Adam[3], M. A. Aldera[6], E. F. Bauermeister[3], R. S. Booth[7], A. G. Botha[3], D. H. Botha[8], L. R. S. Brederode[3,17], Z. B. Brits[3], S. J. Buchner[3], J. P. Burger[3], J. M. Chalmers[3], T. Cheetham[3], D. de Villiers[9], M. A. Dikgale-Mahlakoana[3], L. J. du Toit[8], S. W. P. Esterhuyse[3], B. L. Fanaroff[3], A. R. Foley[3], D. J. Fourie[3], R. R. G. Gamatham[3], S. Goedhart[3], S. Gounden[3], M. J. Hlakola[3], C. J. Hoek[3], A. Hokwana[3], D. M. Horn[3], J. M. G. Horrell[10], B. Hugo[3,2], A. R. Isaacson[3], J. L. Jonas[2,3], J. D. B. L. Jordaan[3,8], A. F. Joubert[3], G. I. G. Józsa[3,2], R. P. M. Julie[3], F. B. Kapp[3], J. S. Kenyon[2], P. P. A. Kotzé[3], H. Kriel[3], T. W. Kusel[3], R. Lehmensiek[8,11], D. Liebenberg[3], A. Loots[12], R. T. Lord[3], B. M. Lunsky[3], P. S. Macfarlane[3], L. G. Magnus[3], C. M. Magozore[3], O. Mahgoub[3], J. P. L. Main[3], J. A. Malan[3], R. D. Malgas[3], J. R. Manley[3], M. D. J. Maree[3], B. Merry[3], R. Millenaar[3], N. Mnyandu[3], I. P. T. Moeng[3], T. E. Monama[3], M. C. Mphego[3], W. S. New[3], B. Ngcebetsha[3,2], N. Oozeer[3,13], A. J. Otto[3], S. S. Passmoor[3], A. A. Patel[3], A. Peens-Hough[3], S. J. Perkins[3], S. M. Ratcliffe[3], R. Renil[3], A. Rust[3], S. Salie[3], L. C. Schwardt[3], M. Serylak[3,14], R. Siebrits[3], S. K. Sirothia[3,2], O. M. Smirnov[2,3], L. Sofeya[3], P. S. Swart[3], C. Tasse[15,2], D. T. Taylor[3], I. P. Theron[8,2], K. Thorat[3,2], A. J. Tiplady[3], S. Tshongweni[3], T. J. van Balla[3], A. van der Byl[3], C. van der Merwe[3], C. L. van Dyk[16], R. Van Rooyen[3], V. Van Tonder[3], R. Van Wyk[3], B. H. Wallace[3], M. G. Welz[3] & L. P. Williams[3]

*1. Department of Physics, University of Oxford, Oxford, UK.*
*2. Department of Physics and Electronics, Rhodes University, Grahamstown, South Africa.*
*3. South African Radio Astronomy Observatory, Cape Town, South Africa.*
*4. National Radio Astronomy Observatory, Charlottesville, VA, USA.*
*5. CIERA and Department of Physics and Astronomy, Northwestern University, Evanston, IL, USA.*
*6. Tellumat (Pty) Ltd, Retreat, South Africa.*
*7. Chalmers University of Technology, Gothenburg, Sweden.*
*8. EMSS Antennas (Pty) Ltd, Stellenbosch, South Africa.*
*9. Department of Electrical and Electronic Engineering, Stellenbosch University, Stellenbosch, South Africa.*
*10. IDIA, University of Cape Town, Rondebosch, South Africa.*
*11. Department of Electrical Engineering, Cape Peninsula University of Technology, Bellville, South Africa.*
*12. Presidential Infrastructure Coordinating Commission, Pretoria, South Africa.*
*13. African Institute for Mathematical Sciences, Muizenberg, South Africa.*
*14. Department of Physics and Astronomy, University of the Western Cape, Bellville, South Africa.*
*15. GEPI, CNRS, PSL Research University, Meudon, France.*
*16. Peralex Electronics (Pty) Ltd, Bergvliet, South Africa.*
*17. Present address: SKA Organisation, Jodrell Bank, UK.*
*\*e-mail:* ian.heywood@physics.ox.ac.uk; fernando@ska.ac.za

**The Galactic Centre contains a supermassive black hole with a mass of four million Suns[1] within an environment that differs markedly from that of the Galactic disk. Although the black hole is essentially quiescent in the broader context of active galactic nuclei, X-ray observations have provided evidence for energetic outbursts from its surroundings[2]. Also, although the levels of star formation in the Galactic Centre have been approximately constant over the past few hundred million years, there is evidence of elevated short-duration bursts[3], strongly influenced by interaction of the black hole with the enhanced gas density present within the ring-like central molecular zone[4] at Galactic longitude $|l|$ < 0.7 degrees and latitude**

**|$b$| < 0.2 degrees. The inner 200-parsec region is characterized by large amounts of warm molecular gas[5], a high cosmic ray ionization rate[6], unusual gas chemistry, enhanced synchrotron emission[7,8], and a multitude of radio-emitting magnetised filaments[9], the origin of which has not been established. Here we report radio imaging that reveals a bipolar bubble structure, with an overall span of 1 degree by 3 degrees (140 parsecs × 430 parsecs), extending above and below the Galactic plane and apparently associated with the Galactic Centre. The structure is edge-brightened and bounded, with symmetry implying creation by an energetic event in the Galactic Centre. We estimate the age of the bubbles to be a few million years, with a total energy of $7 \times 10^{52}$ ergs. We postulate that the progenitor event was a major contributor to the increased cosmic-ray density in the Galactic Centre, and is in turn the principal source of the relativistic particles required to power the synchrotron emission of the radio filaments within and in the vicinity of the bubble cavities.**

We observed the Galactic Centre region with the MeerKAT radio telescope[10], resulting in a deep mosaic spanning several square degrees, with a central frequency of 1,284 MHz and 6 arcsec angular resolution (see Methods). Many new radio structures are revealed, the most important of which is the pair of bounded, bipolar bubbles, spanning 430 pc across the Galactic plane, shown in Fig. 1. The radio emission is non-thermal, with spectral-index measurements consistent with synchrotron radiation, with a cooling time of 1–2 Myr (see Methods). The symmetry of the bubbles about the Galactic Centre implies that the progenitor event took place in the vicinity of the strong radio source Sgr A*, known to be coincident with the central black hole.

Coincident ionized gas near the base of the radio bubbles has a velocity dispersion of 30 km s$^{-1}$ with a central velocity of 0 km s$^{-1}$ (ref. [11]). Assuming no substantial deceleration, this yields a dynamical timescale for the bubbles of approximately 7 Myr. On the basis of this age, the progenitor event of the radio bubbles may be associated with the formation of the young nuclear star cluster within 0.5 pc of Sgr A* 6 Myr ago[12]. Although this age is greater than the synchrotron lifetime given above, the acceleration of relativistic electrons is likely from shocks generated in the initiating event. As such, the shocks will continue to propagate and accelerate fresh radiating electrons.

Evidence that bursts of activity from the nuclear black hole or stellar populations can drive energetic bipolar outflows exists in the form of the Fermi bubbles[13]. Identified via γ-ray imaging, these bipolar structures have approximately coincident polarized synchrotron radio emission[14], and extend to 50° away from the Galactic plane. Their estimated age ranges from 1 Myr for leptonic emission models[15] to >1 Gyr for hadronic models[16]. The edge-brightened caps at the extrema of the radio bubbles (Fig. 1) provide morphological evidence for a time-bounded energetic event, as opposed to (or perhaps with contribution from) steady-state outflows that could be collimated perpendicular to the plane by a ring-like structure. However, as with the Fermi bubbles, it remains unclear whether the origin of the radio bubbles is tied to direct black-hole influence (such as through tidal disruption events or fragmentation of an accretion disk) or is related to a starburst process, or possibly some combination of the two.

Previous observations of the region with single-dish radio telescopes have detected the Galactic Centre lobe, a synchrotron-emitting feature extending to 140 pc at northern Galactic latitudes[7,8]. Our data reveal this to be part of the much larger, continuous bipolar radio structure (Fig. 1). The radio bubbles also trace X-ray structures on similar scales, thought to be the result of episodic periods of activity in the vicinity of Sgr A*[17,18]. The correspondence between radio and X-ray emission is particularly remarkable at southern Galactic latitudes, where the radio bubble appears to almost precisely bound the ionised plasma revealed by enhanced X-ray emission (Fig. 2).

Observations of $H_3^+$ absorption lines imply a high cosmic-ray energy density ($\epsilon_{CR}$) within the central molecular zone[6], approximately two orders of magnitude higher than in the Galactic disk[19]. It is reasonable to assume that this has a latitudinal decline away from the plane, consistent with the thermal pressure decline seen in X-rays[18]. For a cylindrical volume of $4 \times 10^{62}$ cm$^3$ and an average $\epsilon_{CR} = 10$ eV cm$^{-3}$ throughout the bubbles (see Methods), we derive a total cosmic ray energy within the cavities of $7 \times 10^{51}$ erg. This is comparable to the thermal energy derived from X-ray measurements[18] of $4 \times 10^{52}$ erg. These observations are consistent with modelling of X-ray and radio synchrotron emission towards the Galactic Centre, which indicates that similar cosmic ray and gas pressures are required to drive outflows along the in-situ magnetic field lines[20]. Assuming that 10% of the energy emerges in the form of cosmic rays[21], we derive a total energy budget for the progenitor event of $7 \times 10^{52}$ erg.

Whatever the mechanism by which the progenitor event occurred, the creation of a large population of relativistic cosmic ray particles is powering the radio synchrotron emission of the bubbles. The same population could thus also be driving the emission from the magnetized radio filaments, of which more than 100 have been identified in the Galactic Centre region, and nowhere else, since their discovery 35 years ago[22] (Fig. 2). The filaments are linearly polarized and have synchrotron spectra, their highly linear morphology tracing locally or globally ordered magnetic fields[23] with strengths of the order of[24] 100 µG. To date there has been no conclusive explanation for their origin[25,26]. One of the key unknowns is the mechanism by which particles are injected and accelerated to relativistic speeds in order to generate the observed synchrotron emission around the magnetic fields. Comparing the distribution and brightness of the filaments to the boundary of the radio bubbles reveals a clear spatial association between the two (Fig. 2). This indicates a causal connection between the formation of magnetised filaments and the radio bubbles. As viewed in projection, the bubble cavities contain a significantly enhanced number of bright and multiple-filament complexes, indicating that the event that generated the bubbles was also the source of the relativistic particles required to illuminate the filaments on a large scale. The number of filaments is also seen to decline latitudinally, consistent with a corresponding decline in $\epsilon_{CR}$. Filaments observed beyond the radio bubble boundaries could be illuminated by the same event if, by analogy with supernova remnants[21], the radio emitting boundary of the bubbles lags the expanding cosmic ray front.

The 'radio arc'[9] at $l = 0.2°$, transversely resolved into a complex of numerous magnetized filaments, is revealed to be coincident with the corresponding edge of the radio bubbles, above and below the plane (Fig. 2). The filaments that make up the radio arc are also revealed to be far longer than previously known, with an extent seen to be at least 1.5° (210 pc) before fading below detectability. There is also a corresponding western counterpart to the radio arc, coincident with the Sgr C complex at $l = 359.4°$. Although the peak flux density of the counterpart (32 mJy per beam) is close to that of the radio arc (27 mJy per beam) in the mid-plane region, the radio arc consists of far more filaments of comparable brightness. The radio arc complex also remains brighter away from the plane, suggesting that the expanding bubbles have a stronger shock interaction at positive Galactic longitudes, a direction in which the inner regions of the Milky Way are known to have a density enhancement[4].

We may be witnessing a less energetic version of a process similar to that which created the Fermi bubbles and their corresponding radio emission. The typically inferred[14,27] value of $10^{56}$ erg for the total energy content of the Fermi bubbles significantly exceeds that of the radio bubbles. However, the radio bubbles may be one example of a series of such intermittent events, possibly combined with weaker, steadier outflows[28], with the cumulative influence of these events being responsible for the observed radio, X-ray and **γ**-ray structures that connect the Galactic Centre to the emission observed at higher latitudes.

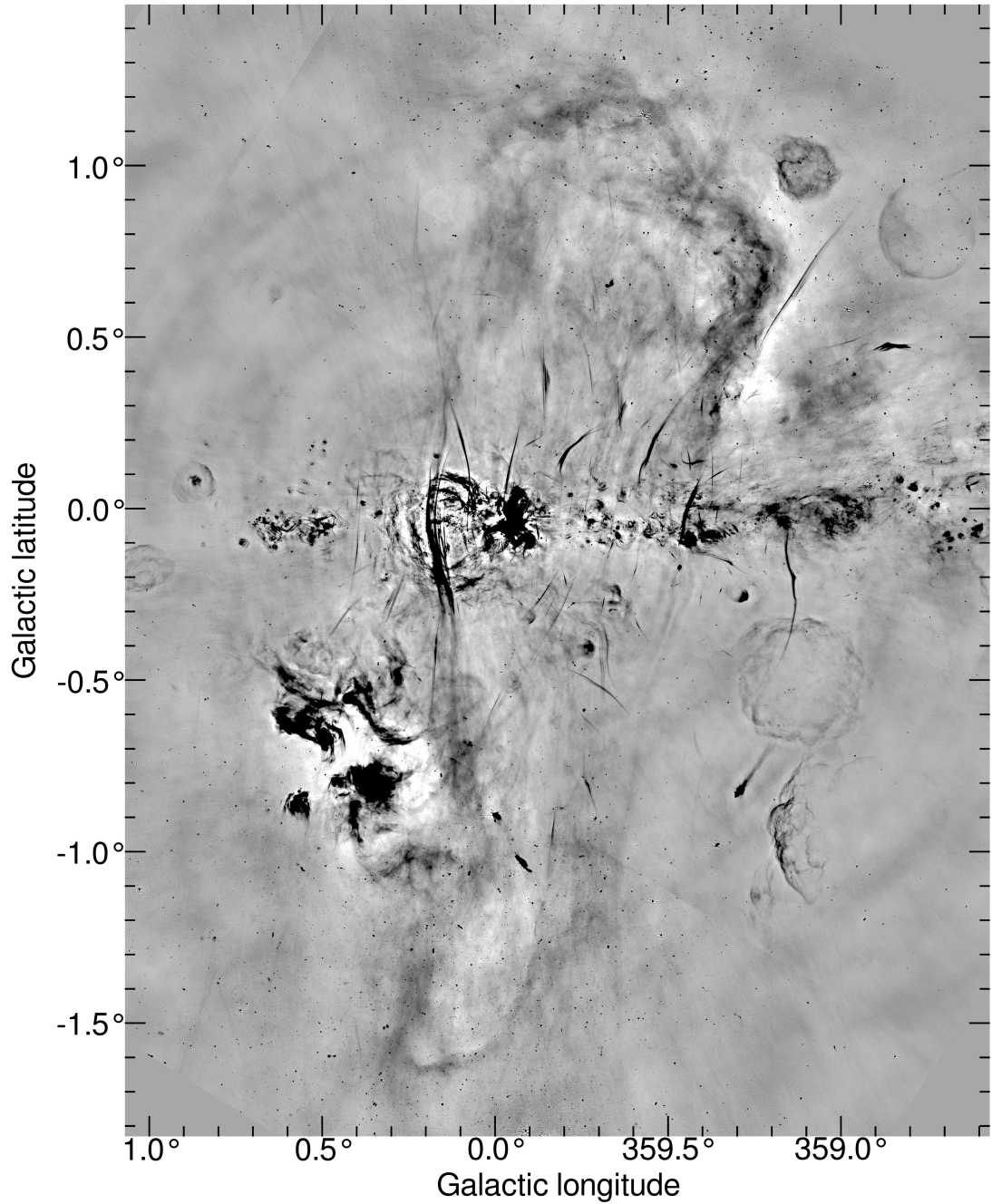

**Figure 1: Radio emission from the Galactic Centre bubbles.** An inverted greyscale radio image of the Galactic Centre region at 1,284 MHz, showing a bounded pair of synchrotron-emitting edge-brightened radio bubbles. The morphology and symmetry about the Galactic Centre strongly suggest creation by an energetic event in the vicinity of Sgr A*. The bubbles have since expanded to span Galactic latitudes, $b$, of −1.6° to +1.3°, corresponding to a total major axis length of 430 pc, and dwarfing any other coherent radio structure visible in this image. The bubbles cross the Galactic plane between longitudes, $l$, of −0.6° and +0.2°, for an axial ratio of 3.6. The bubble structure at $b = 0.0°$ is offset from Sgr A* itself by 20 pc towards negative longitudes, in which direction the density of the surrounding material is known to be lower[4]. The numerous compact sources away from the Galactic plane are predominantly background galaxies. The Galactic Centre region has very strong radio emission on large angular scales. This causes the distinct white regions around the foreground H II region (Fig. 2) and the western edge of the northern bubble, which are artefacts resulting from a fundamental upper limit to the angular scales that can be detected by radio interferometers. This is an important consideration to bear in mind when comparing interferometer maps such as this one to those obtained with single-dish radio telescopes.

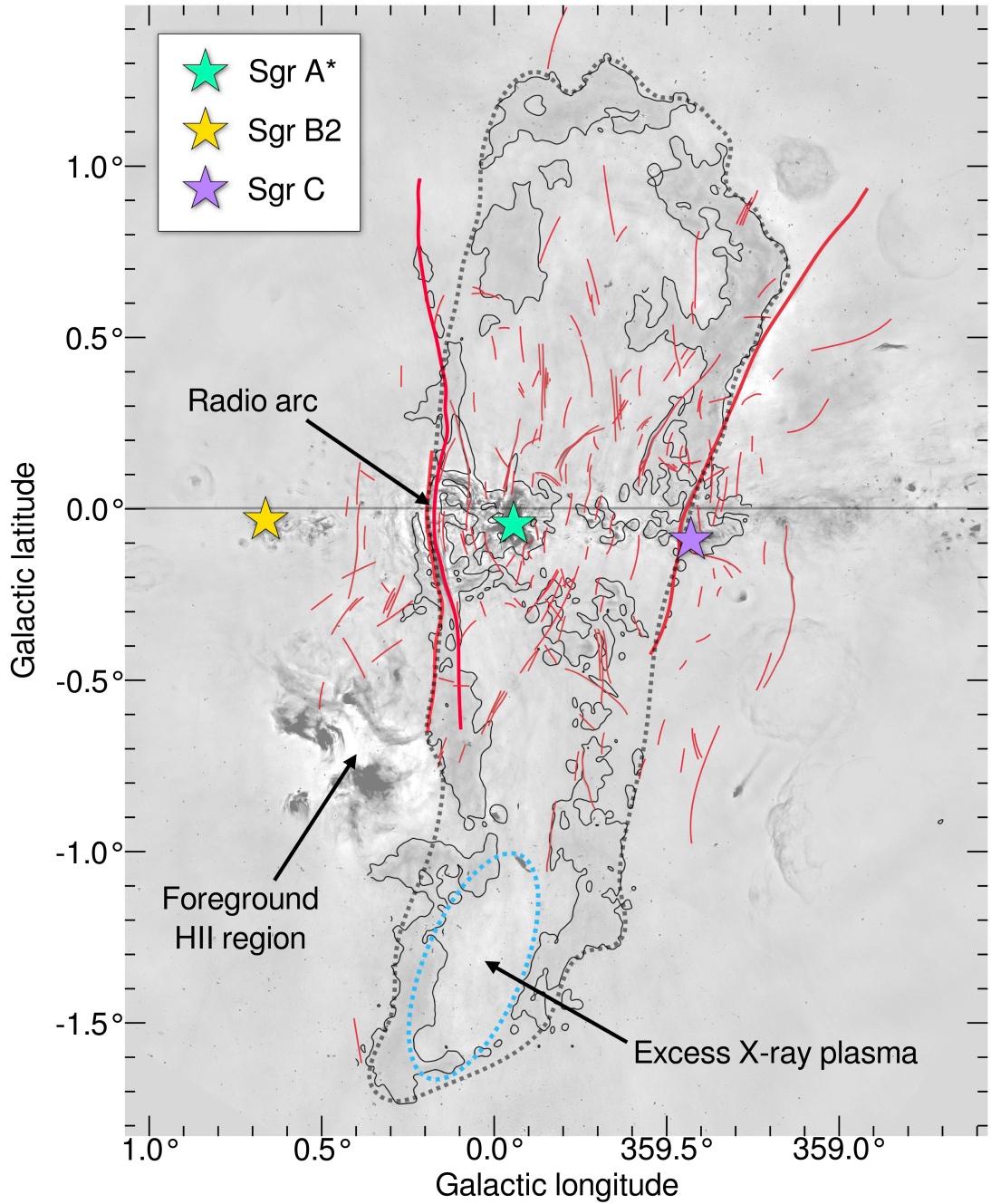

**Figure 2: Major features of the Galactic Centre radio bubbles.** The figure shows a replication of the region shown in Fig. 1. The dashed grey line shows the outline of the radio bubbles (as traced over an isophote of the relevant region of the background radio image convolved to arcminute resolution) in order to guide the eye for comparison to other relevant features. The location of the strong radio source Sgr A* is marked, along with the positions of the giant molecular cloud complexes Sgr B2 and Sgr C (see key). The location of a bright foreground H II is noted so as not to be confused with the emission of the radio bubbles. The most prominent magnetized filaments are marked in red, including the longest filaments making up the 'radio arc' structure, and the corresponding counterpart near Sgr C, that along with the radio arc are coincident with the longitudinal boundaries of the radio bubbles. Note that many of the red marks represent complexes consisting of multiple parallel filaments. The decline in the number density of radio filaments away from the plane supports the picture of declining cosmic-ray energy density with Galactic latitude. The location of enhanced X-ray emission within the southern tip of the bubbles[17,18] is also marked (dashed blue line), and examination of this region in Fig. 1 shows almost perfect correspondence between the elevated X-ray counts and a cavity in the radio emission, consistent with the bubble being supported internally by the excess X-ray emitting plasma as marked.

**METHODS**

**Observations and data reduction.** The Galactic Centre region was observed by MeerKAT between 6 May 2018 and 2 July 2018. A total of 26 pointings were observed, using a minimum of 60 and a maximum of 64 antennas for each observation, providing up to 2,016 baselines of maximum length 8 km. The minimum separation between antenna pairs in the array is 29 m, so at the band centre frequency of 1,284 MHz, a zenith observation with MeerKAT would be insensitive to structures larger than 0.46°. Individual 10-h tracks were typical, including observations of required calibrator sources. The correlator was configured for an 8-s integration time per visibility point, with 4,096 frequency channels covering the full MeerKAT L-band (nominally 856 MHz wide, but with a usable band of 900–1,670 MHz).

The data were converted from the native MVFv4 format to a CASA Measurement Set using the KAT Data Access Library (katdal) package, averaging by a factor of 4 in frequency to reduce the number of channels to 1,024. Subsequent flagging and reference calibration steps were performed using the CASA package. Visibilities with amplitudes of exactly zero were discarded. An automatic algorithm for the identification and removal of radio frequency interference was then used, namely the rflag algorithm within CASA's flagdata task. This was run on both target and calibrator observations.

Delay and bandpass corrections were derived from the observations of the primary calibrator source PKS B1934−638, which was also used to set the absolute flux scale. Time-dependent complex gains for each antenna were derived using the observations of the secondary calibrator source 1827−360. This was chosen for its brightness (8 Jy at 1,284 MHz), and visited frequently (a 1-min integration for every 10-min of target data), reasoning that amplitude self-calibration of the target field would be difficult or impossible due to the bright, morphologically complex structures. Following this, the instrumental corrections were all applied to the relevant target field.

The first pass of imaging of the calibrated data for each target field was performed using wsclean[29]. Multiscale deconvolution[30] and automasking were enabled, with the deconvolution process terminating after 250,000 clean iterations. In order to reduce the sidelobes of the point spread function, the gridded visibilities were weighted using the Briggs method, with a robustness parameter of −0.3. This results in an angular resolution of 6 arcsec. Each image is 12,000 × 12,000 pixels of 1.5 × 1.5 arcsec$^2$, for a total field of view of 5° × 5°. Although the field of view of MeerKAT corresponding to the high-gain region of the antenna primary beam at 1,284 MHz is approximately 1° across, a large image is required to deconvolve and model structures that are detected (albeit attenuated) through the sidelobes of the antenna primary beam. This attenuation is an effect that we capitalise on when forming the images presented in this paper, with further details given below. To model the apparent spectral variations of the sky across the band, deconvolution was performed in eight sub-bands. In wsclean's implementation of multifrequency synthesis, peak finding during the minor cycle occurs in the image formed from the entire bandwidth in order to maximise the signal-to-noise ratio, and then each of the sub-bands is deconvolved independently.

Following the creation of the first deconvolved image, a threshold mask was manually created down to the brightness level that ruled out any spurious features. The imaging process was then repeated, with the second pass of deconvolution restricted to the region defined by the mask in order to generate an artefact-free model for self-calibration purposes. The frequency-dependent sky model derived from the deconvolution process was inverted into a set of model visibilities by wsclean at the end of the imaging process.

Phase-only antenna-based gain corrections were derived from the resulting model visibilities using the CubiCal software[31]. Solutions were derived for every 128 s of data, with independent solutions derived for four 256-channel groups. Spacings shorter than 300 m were excluded from the calibration process, as the large-scale emission on the shortest spacings is difficult to model using clean-based deconvolution techniques. The self-calibration solutions were applied, and the third and final imaging operation then took place on the corrected data using the same set-up as the second step.

Of the 26 pointings observed, only four were used to make the images presented in this Letter. These were placed at comparatively high Galactic latitudes, with coordinates provided in Extended Data Table 1. Pointings for which the bright Sgr A* complex is close to the high-gain region of the primary beam are dynamic-range-limited to the point where the faint, diffuse emission of the radio bubbles is confused by residual sidelobe structures. As such we only make use of a subset of four pointings to make the image used in Figs. 1 and 2, for which Sgr A* is attenuated enough by the antenna primary-beam response that the bubbles can be clearly detected. Sgr A* and many of the radio structures in the Galactic plane are bright enough to still be visible through the primary beam sidelobes, which are stable enough for the deconvolution to be effective. A simple linear mosaic of these four pointings was generated using the Montage software. Since no primary beam correction has been applied, the images are not presented with flux density scale bars, and Figs. 1 and 2 are not suitable for absolute photometric studies. The two flux density values quoted (for the radio arc and corresponding counterpart) are measured from the primary-beam corrected images of the two pointings of the 26 that are closest to the relevant structures (GC31 and GC22 in Extended Data Table 1).

**Spectral index measurements of the southern cap.** In order to confirm the non-thermal (synchrotron) nature of the radio emission from the bubbles, we formed spectral index maps by using the eight sub-band images for pointing GCXS30 (Extended Data Table 1), which targets almost precisely the southern cap of the radio bubbles. Spectral index (α) is defined via the power law:

$$S(\nu) \propto \nu^{\alpha} \qquad (1)$$

where $S$ is the flux density and $\nu$ is the frequency. Each of the sub-band images has 107 MHz of bandwidth. Not all eight were used, as for some of the lower sub-bands Sgr A* entered the main lobe of the antenna primary beam and limited the performance of the broader image. Some sub-bands also covered the parts of the MeerKAT L-band spectrum that suffer from radio frequency interference. For GCXS30 the first, second and fourth sub-bands were not used (see Extended Data Table 2).

Each of the sub-band images was convolved to an angular resolution of 30 arcsec using CASA's ia.convolve2d tool. The images were primary-beam corrected by dividing them by a frequency-dependent model of the MeerKAT primary beam. The convolved and primary-beam corrected images were then placed into a three-dimensional image cube. Following this, a spectral index map was generated by fitting for α by extracting the spectrum of each spatial pixel in the cube. Performing a power-law fit across five spectral points in this way is more robust to the effects of noise or imaging artefacts compared to simply dividing the band into two, however there is an associated sensitivity penalty. To ensure that we only fit for spectral indices in regions of high signal-to-noise ratio, a pixel mask was applied to the spatial axes of the cube that excluded emission below a certain brightness threshold determined by eye from the highest-frequency plane, that is, the one with the lowest signal-to-noise ratio. For GCXS30 the threshold was 0.14 mJy per beam.

An example of one of the sub-band images (central frequency 1,123 MHz) is shown in Extended Data Fig. 1 for pointing GCXS30, corresponding to the southernmost tip of the radio bubbles. The masked spectral-index map is overlaid with the corresponding colour scale. Marked on this figure are the mean spectral-index measurements for two regions of the radio bubbles, and the 1σ error estimate taken to be the standard deviation of spectral-index pixel values, weighted by their corresponding total intensity values. As a consistency check, five compact sources in the map were also examined, as annotated on Extended Data Fig. 1, and with measured spectral-index values listed in Extended Data Table 3. The measured spectral indices are a mixture of values consistent with synchrotron emission (sources A, B, and E; likely to be radio emission from typical background galaxies external to our own) and flatter spectrum sources (C and D). Only source D has a catalogued counterpart that potentially confirms its nature, namely a planetary nebula[32]. The measured spectral index is consistent with the typical radio spectrum of planetary nebulae, namely a free-free dominated emission process that is self-absorbed at MeerKAT L-band frequencies[33].

Spectral-index measurements for the northern cap of the bubbles using pointing GCXN17 proved difficult because of residual sidelobe emission associated with Sgr A*. The radial position of the residual sidelobes is a function of frequency, leading to frequency-dependent brightness corruptions in the sub-band images. In any case, considering our key finding that the bubbles represent a single coherent structure, we can instead draw on the spectral-index measurements of the Galactic Centre lobe previously made with single-dish telescopes[8] across a larger bandwidth than the MeerKAT observations provide. These also confirm the synchrotron nature of the emission, consistent with our findings at the southern cap.

**Synchrotron cooling time.** Single-dish telescope observations reveal the spectrum of the large-scale radio emission from the inner 2° × 1° around the Galactic Centre to be well characterized by a broken power law[34], transitioning from $\alpha = -0.25$ to $\alpha = -1.1$ at a break frequency of approximately 3,300 MHz. This is consistent with earlier studies that estimated $\alpha = -0.7$ over the inner 6° × 2° region[35]. The estimated equipartition magnetic field from these studies is between 10 μG and 20 μG.

Assuming that the spectral index (equation (1)) of synchrotron emission averaged over the bubble is $\alpha = -0.5$ between frequencies of $10^7$ Hz and $10^{12}$ Hz, then the synchrotron lifetime (which is approximately the age of the source, unless electrons are re-accelerated) can be calculated[36] to be 1.9 Myr. This assumes a magnetic field strength of 20 μG, and equipartition of energies between the magnetic field and the cosmic ray particles. A flatter average radio spectral index of −0.2 results in a synchrotron lifetime of 1 Myr.

**Gradient of the cosmic-ray energy density.** The average cosmic-ray energy density within the bubble volume can be quantitatively estimated using measurements of the average radio synchrotron brightness over relevant regions. For this we make use of 330-MHz data[35], combining observations from both an interferometer (the Very Large Array) and a single-dish telescope (the Green Bank Telescope), in order to be sensitive to large angular-scale emission and image the region at suitable angular resolution.

Ionization rates in the Galactic Centre are approximately two orders of magnitude higher than those in the Galactic disk[6,37]. Under the reasonable assumption that cosmic-ray energy densities $\epsilon_{CR}$ scale linearly with ionization rate, we can scale the local value[38] of $\epsilon_{CR}$ = 1.8 eV cm$^{-3}$ to conservatively estimate a value of 100 eV cm$^{-3}$ in the Galactic Centre region, close to the plane and within the central molecular zone. The flux density of synchrotron radiation $S$ is dependent on the product of the magnetic field $B^{3/2}$ and the density of relativistic electrons, the latter being related directly to $\epsilon_{CR}$. Thus, the ratio of the radio flux density measurements between two regions of interest can be used to scale the value of $\epsilon_{CR}$ from one region to another, under the assumption that the magnetic field remains approximately constant over both. The flux density ratio of the bubble cavities (regions 1° above and below the plane) to that of the central Galactic plane is determined to be (3.85 mJy per beam / 37.5 mJy per beam) = 0.103, using the aforementioned radio image[35]. This suggests that our adopted average value of $\epsilon_{CR}$ = 10 eV cm$^{-3}$ throughout the bubble volume is justified.

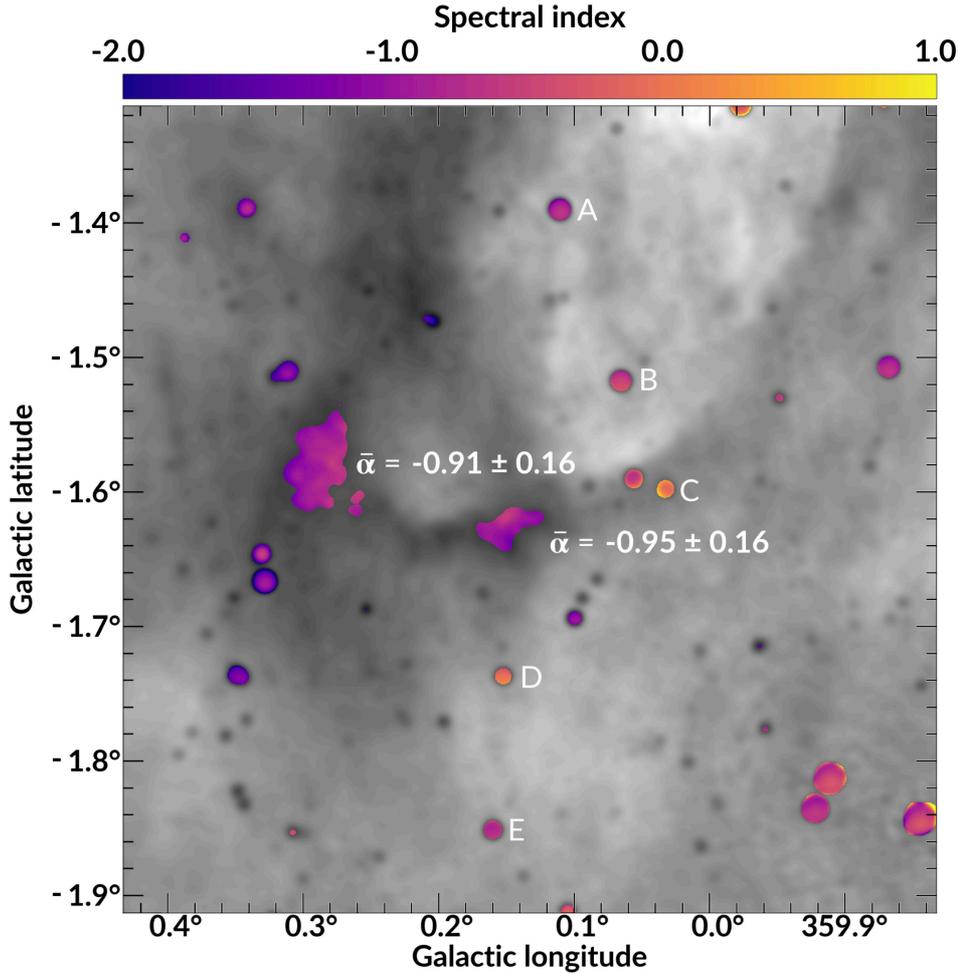

**Extended Data Figure 1: Spectral-index map of the region surrounding the southern tip of the radio bubbles.** A single sub-band image at 1,123 MHz from pointing GCXS30 (see Extended Data Table 1), convolved to 30-arcsec resolution and corrected for primary-beam attenuation effects, is shown in greyscale. Overlaid upon this is a map of the spectral index, derived by fitting for this quantity through an image cube formed from five such sub-band images (see Extended Data Table 2). The spectral index values for features that have a high signal-to-noise ratio detection in each sub-band can be reliably measured using this method, with values as indicated by the colour bar. The mean and 1σ spectral-index values for two such extended regions are marked on the map (in the format ᾱ = mean ± 1σ) and are consistent with non-thermal synchrotron emission. The spectral indices of the additional compact sources marked A–E are also measured for verification purposes, with the results listed in Extended Data Table 3.

| ID | *l* (deg) | *b* (deg) |
|---|---|---|
| GCX30 | 0.0843 | −1.1912 |
| GCXS30 | 0.1328 | −1.6127 |
| GCX17 | 359.6643 | 1.0819 |
| GCXN17 | 359.7459 | 1.6784 |
| GC31 | 0.2699 | −0.1684 |
| GC22 | 359.5486 | −0.1834 |

**Extended Data Table 1: IDs and pointing centres for the MeerKAT observations.** The first four pointings are relevant for Figs. 1 and 2, and for the spectral-index measurements described in Methods. The last two pointings are used to extract the peak flux density measurements for the radio arc and the counterpart bounding filament structure close to the Sgr C region. Positions are given in Galactic coordinates.

| Sub-band # | Centre frequency (MHz) | Used? |
|---|---|---|
| 1 | 909.4 | N |
| 2 | 1016.4 | N |
| 3 | 1123.4 | Y |
| 4 | 1230.4 | N |
| 5 | 1337.4 | Y |
| 6 | 1444.4 | Y |
| 7 | 1551.4 | Y |
| 8 | 1658.4 | Y |

**Extended Data Table 2: Central frequencies of the eight sub-band images.** The third column indicates (N, no; Y, yes) whether that sub-band was included in the production of the spectral-index map in Extended Data Fig. 1. See Methods for details.

| Field source | Spectral index |
|---|---|
| A | −0.85 ± 0.17 |
| B | −0.55 ± 0.15 |
| C | 0.12 ± 0.23 |
| D | −0.07 ± 0.14 |
| E | −0.73 ± 0.08 |

**Extended Data Table 3: Additional spectral-index measurements for verification purposes.** The entries A–E correspond to the compact sources marked accordingly on Extended Data Fig. 1. The listed uncertainties represent 1σ confidence levels. See Methods for details.

**Data availability** The data that support the findings of this study are available from a corresponding author upon reasonable request.

**Acknowledgements** The MeerKAT telescope is operated by the South African Radio Astronomy Observatory, which is a facility of the National Research Foundation, an agency of the Department of Science and Innovation. We acknowledge use of the Inter-University Institute for Data Intensive Astronomy (IDIA) data intensive research cloud for data processing. IDIA is a South African university partnership involving the University of Cape Town, the University of Pretoria and the University of the Western Cape. This research made use of Montage, which is funded by the US National Science Foundation under grant number ACI-1440620, and was previously funded by the National Aeronautics and Space Administration's Earth Science Technology Office, Computation Technologies Project, under cooperative agreement number NCC5-626 between NASA and the California Institute of Technology. I.H. acknowledges support from the Oxford Hintze Centre for Astrophysical Surveys which is funded through generous support from the Hintze Family Charitable Foundation. F.Y-Z. is partially supported by the grant AST-0807400 from the US National Science Foundation.

**Author contributions** I.H. and F.C. planned the MeerKAT observations presented here. I.H. performed the calibration and imaging of the observations. I.H. wrote the manuscript together with F.Y-Z., F.C. and W.D.C. All other authors have contributed to one or more of the planning, design, construction, commissioning or operation of the MeerKAT radio telescope.

**Correspondence and requests for materials** should be addressed to I.H. or F.C.